\documentclass[aps,prl,amsmath,amssymb,floatfix,twocolumn,amsmath,superscriptaddress,twocolumn,nofootinbib,tighten,letterpaper]{revtex4}
\usepackage{multirow}
\usepackage{subfigure}
\usepackage{color}
\usepackage{mathrsfs}
\usepackage{hyperref}
\usepackage[normalem]{ulem}
\usepackage{bm}

\usepackage{amssymb}   
\usepackage{amsmath}
\renewcommand\vec[1]{\ensuremath\boldsymbol{#1}} 

\usepackage{amsfonts, relsize, color}
\usepackage{graphics}
\usepackage{graphicx}
\usepackage{subfigure}
\usepackage{hyperref}
\usepackage{color}
\usepackage{comment}

\newcommand{\bra}[1]{\left< #1 \right|}
\newcommand{\ket}[1]{\left| #1 \right>}

\begin{document}

\title{Topological insulators on fractal lattices: A general principle of construction}

\author{Daniel J.\ Salib}
\affiliation{Department of Physics, Lehigh University, Bethlehem, Pennsylvania, 18015, USA}

\author{Aiden J.\ Mains}
\affiliation{Department of Physics, Lehigh University, Bethlehem, Pennsylvania, 18015, USA}

\author{Bitan Roy}
\affiliation{Department of Physics, Lehigh University, Bethlehem, Pennsylvania, 18015, USA}

\date{\today}

\begin{abstract}
Fractal lattices, featuring the self-similarity symmetry, are often geometric descents of parent crystals, possessing all their discrete symmetries (such as rotations and reflections) except the translational ones. Here, we formulate three different general approaches to construct real space Hamiltonian on a fractal lattice starting from the Bloch Hamiltonian on the parent crystal, fostering for example strong and crystalline topological insulators resulting from the interplay between the nontrivial geometry of the underlying electronic wave functions and the crystal symmetries. As a demonstrative example, we consider a generalized square lattice Chern insulator model and within the framework of all three methods we successfully showcase incarnations of strong and crystalline Chern insulators on the Sierpi\'nski carpet fractal lattices. The proposed theoretical framework thus lays a generic foundation to build a tower of topological phases on the landscape of fractal lattices.   
\end{abstract}

\maketitle

\begin{figure}[t!]
\includegraphics[width=0.95\linewidth]{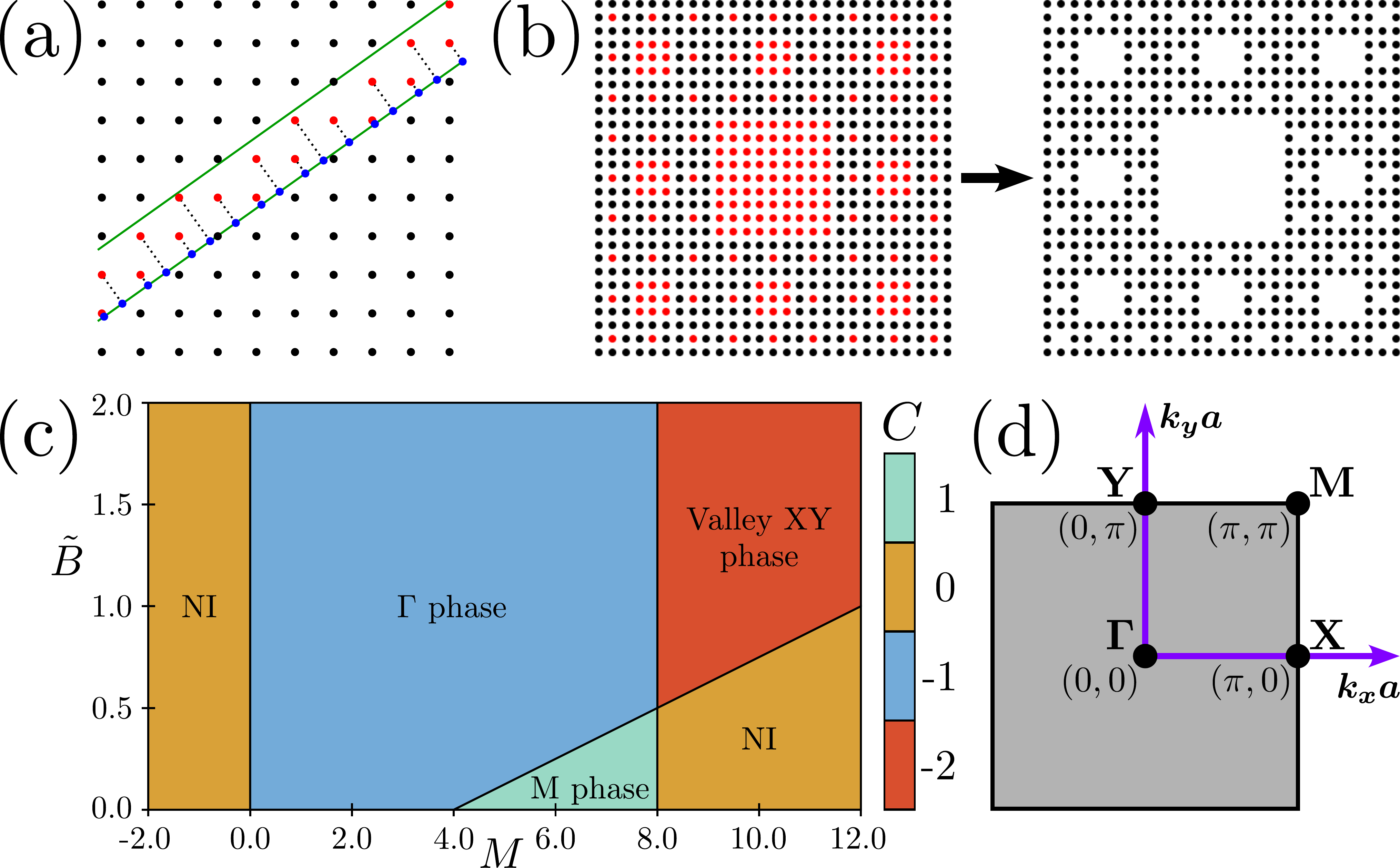}
\caption{Construction of (a) a one-dimensional (1D) Fibonacci quasicrystal (blue dots), realized by projecting the red sites of the parent square lattice, living within the green lines, each with a slope $(1-\sqrt{5})/2$, on a 1D chain, and (b) a two-dimensional (2D) Sierpi\'nski carpet fractal lattice from a parent square lattice, built by eliminating its red colored sites. (c) Phase diagram of the generalized Qi-Wu-Zhang model for $t_1=t_2=B=1$ [Eqs.~\eqref{eq:QWZ} and~\eqref{eq:dvec}], fostering strong topological insulators with the band inversion at the $\Gamma$ ($\Gamma$ phase) and ${\rm M}$ (${\rm M}$ phase) points as well as a crystalline topological insulator with the band inversion simultaneously at the ${\rm X}$ and ${\rm Y}$ points (valley ${\rm XY}$ phase)~\cite{CTITh3}. Each phase is identified by a distinct quantized Chern number $C$ (see color bar) and $C=0$ for the normal insulator (NI) [Eq.~\eqref{eq:chernnumber}]. We arrive at an identical phase diagram in terms of the Bott index [Eq.~\eqref{eq:bottindex}]. (d) 2D Brillouin zone of a square lattice with lattice spacing $a$, showing the $\Gamma$, ${\rm M}$, ${\rm X}$, and ${\rm Y}$ points. Along the $\tilde{B}=0$ line of the phase diagram we also set $t_2=0$ to switch off all hopping processes along the diagonal directions.           
}~\label{fig:Fig1}
\end{figure}

\emph{Introduction}.~Quasicrystals and fractals are prominent members of the structurally diverse family of solids, typically encompassing crystals. As such, quasicrystals are constituted by a set of sites living on a brane inside a higher-dimensional crystal, as shown in Fig.~\ref{fig:Fig1}(a) in terms of the one-dimensional Fibonacci quasicrystal residing within a two-dimensional (2D) square lattice (SL)~\cite{quasicrystal:book}. By contrast, fractals can be built by eliminating specific sites of crystals, such that the resulting structures feature the self-similarity symmetry. This procedure is shown in Fig.~\ref{fig:Fig1}(b) for the Sierpi\'nski carpet fractal lattice (SCFL), emerging out of a SL~\cite{fractal:book}. Thus, together they constitute the geometric descent family of crystals.   

Such geometric correspondences, when imposed on the Hilbert space of the topological Bloch Hamiltonian for a crystal, raise a fascinating possibility of harnessing novel topological phases of matter on quasicrystals~\cite{panigrahi-roy-juricic, tyner-juricic} and fractal lattices, resulting from the intriguing interplay between the geometry of the underlying electronic wavefunctions and the crystal symmetries. Among the plethora of possibilities, strong~\cite{TITh1, TITh2, TITh3, TITh4, TITh5, TITh6, TITh7, TITh8, TITh9, TITh10, TITh11, TITh12} and crystalline~\cite{CTITh1, CTITh2, CTITh3, CTITh4} topological insulators (TIs), about which more in a moment, are the most prominent and commonly occurring ones in quantum crystals that are routinely discovered in nature following the prescriptions of topological quantum chemistry~\cite{TQC1, TQC2, TQC3, TQC4, TQC5, TQC6, TQC7}. Identifying such phases on fractal lattices is the central theme of the current pursuit.

As demonstrative examples, here we showcase the appearance of both strong and crystalline TIs on SCFLs, possessing all the discrete symmetries of a SL, such as the four-fold rotation about the $z$ direction and the reflections about $x$ and $y$ axes, except for the translational ones, starting from a generalized SL model for quantum anomalous Hall or Chern insulators [Figs.~\ref{fig:Fig1}(c) and (d)]. We formulate three different approaches to construct the effective real space Hamiltonian on SCFLs, each of which allows strong and crystalline analogues of the SL Chern insulator. In all these cases, the bulk-boundary correspondence between a nontrivial bulk topological invariant and the resulting edge modes remain operative. Our theoretical formulation, therefore, opens an unexplored territory of exotic topological phases, already cataloged for topological quantum crystals, realizable on their geometric descent fractal lattices.      

\begin{figure}[t!]
\includegraphics[width=0.90\linewidth]{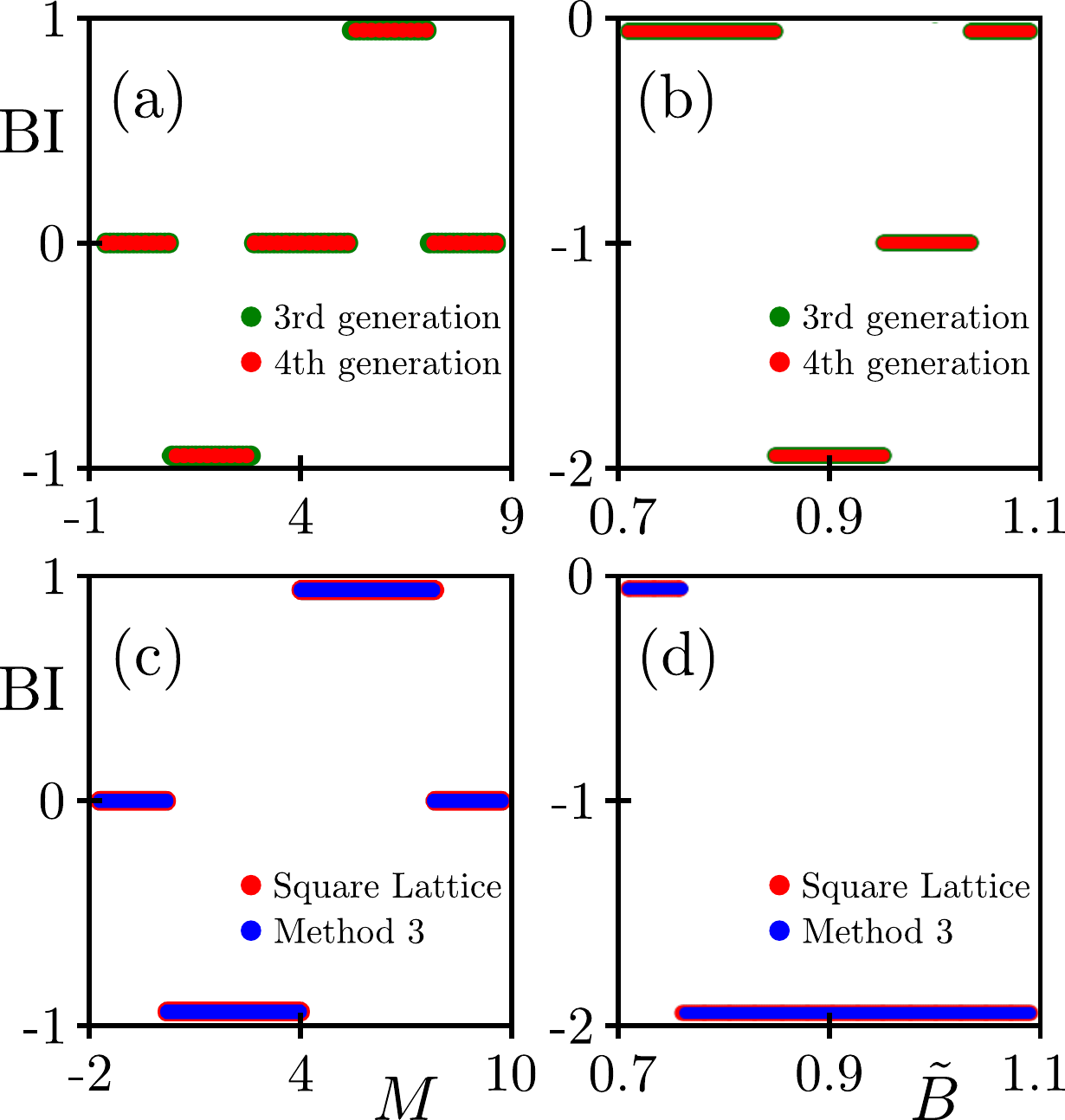}
\caption{Phase diagram on a Sierpi\'nski carpet fractal lattice (SCFL) in terms of the Bott index (BI), see Eq.~\eqref{eq:bottindex}, obtained via method 1 (method of symmetry) and method 2 (method of site elimination) for $t_1=B=1$ and (a) $t_2=\tilde{B}=0$, showing the analogues of the $\Gamma$ phase with ${\rm BI}=-1$ and the ${\rm M}$ phase with ${\rm BI}=+1$, and (b) $t_2=1$ and $M=10$, supporting an analog of the ``valley XY phase" with ${\rm BI}=-2$, besides the normal insulators with ${\rm BI}=0$. These two methods yield identical phase diagrams on SCFL of third and fourth generations, containing 512 and 4096 sites, respectively. In method 1, we set $R_1=(1+\delta)a$ [for (a) and (b)] and $R_2=(\sqrt{2}+\delta)a$ [for (b)], where $\delta \ll 1$ is a small positive number, ensuring that nearest-neighbor and next-nearest-neighbor sites are connected during the numerical analyses. Phase diagram on a parent $L=27$ square lattice and third generation SCFL using method 3 (Method of renormalization) for $t_1=B=1$ and (c) $t_2=\tilde{B}=0$, showing the appearance of the $\Gamma$ (${\rm M}$) phase with ${\rm BI}=-1$ ($+1$), and (d) $t_2=1$ and $M=10$, showing the appearance of the ``valley XY phase" with ${\rm BI}=-2$. The phase diagrams for the SCFL obtained via method 3 are always identical to that of its parent square lattice; see Fig.~\ref{fig:Fig1}(c) and compare with Fig.~S1(c) of the Supplemental Material~\cite{SM}. Note that when $t_2=\tilde{B}=0$, all diagonal hopping is switched off of SL [Eq.~\eqref{eq:dvec}], and we recover the original Qi-Wu-Zhang model with only nearest-neighbor hopping~\cite{QWZ}, yielding the phase diagram on the $\tilde{B}=0$ axis in Fig.~\ref{fig:Fig1}(c) of SL. In (a) and (c), we expose such a line on the fractal lattice.     
}~\label{fig:Fig2}
\end{figure}

\emph{Bloch Hamiltonian}.~The Qi-Wu-Zhang model for quantum anomalous Hall or Chern insulators is given by~\cite{QWZ}
\allowdisplaybreaks[4] 
\begin{equation}~\label{eq:QWZ}
H^{\rm gen}_{\rm QWZ}= \sum_{\vec{k}} \left( c^\dagger_{+,\vec{k}} \:\: c^\dagger_{-,\vec{k}} \right) \: \left[ \sum^{3}_{j=1}\tau_j d_j(\vec{k}) \right] 
\: \left( \begin{array}{c}
c_{+,\vec{k}} \\ 
c_{-,\vec{k}} 
\end{array} \right).
\end{equation}
Fermionic creation (annihilation) operators with parity $\tau=\pm$ and momentum $\vec{k}=(k_x,k_y)$ are $c^\dagger_{\tau,\vec{k}}$ ($c_{\tau,\vec{k}}$). The vector Pauli matrix ${\boldsymbol \tau}=(\tau_1, \tau_2, \tau_3)$ operates on the parity index. The components of $\vec{d}(\vec{k})$ are~\cite{CTITh3} 
\begin{eqnarray}~\label{eq:dvec}
\hspace{-0.5cm} d_1(\vec{k}) &=& t_1 S_x + t_2 C_x S_y, \:\:
d_2(\vec{k}) = t_1 S_y + t_2 C_y S_x, \nonumber \\
\hspace{-0.5cm} d_3(\vec{k}) &=& M-4 B -4 \tilde{B} + 2B \left( C_x + C_y \right) + 4 \tilde{B} C_x C_y,
\end{eqnarray}
where  $S_j=\sin(k_j a)$ and $C_j=\cos(k_j a)$ for $j=x,y$. The hopping amplitude between the orbitals with opposite [same] parities [parity], living on the nearest-neighbor (next-nearest-neighbor) sites of the SL with the lattice spacing $a$ is $t_1$ ($t_2$) [$2B$ ($4 \tilde{B}$)]. The on-site staggered density between two orbitals is $M-4 B -4 \tilde{B} \equiv M_{\rm eff}$. For simplicity, we ignore any particle-hole asymmetry. The corresponding tight-binding Hamiltonian on a SL reads
\begin{eqnarray}~\label{eq:realspaceTB}
\hspace{-0.5cm} && H^{\rm SL}_{\rm TB} =\sum_{\vec{r}} \bigg[ M_{\rm eff} \Psi^{\dagger}_{\vec{r}} \tau_3 \Psi_{\vec{r}}
+ \bigg\{ \bigg( \sum_{j=1,2} \bigg\{ \frac{t_1}{2i} \Psi^{\dagger}_{\vec{r}} \tau_j \Psi_{\vec{r}+\hat{e}_j} \nonumber \\
\hspace{-0.5cm} &&+ B \Psi^{\dagger}_{\vec{r}} \tau_3 \Psi_{\vec{r}+\hat{e}_j} \bigg\} 
+ \frac{t_2}{4i} \sum_{\alpha, \beta=\pm} \Psi^{\dagger}_{\vec{r}+\alpha \hat{e}_1} \left( \beta \tau_1 + \alpha \tau_2 \right) \Psi_{\vec{r}+\beta\hat{e}_2} \nonumber \\
\hspace{-0.5cm} &&+ \tilde{B} \sum_{\alpha, \beta=\pm} \Psi^{\dagger}_{\vec{r}+\alpha \hat{e}_1} \tau_3 \Psi_{\vec{r}+\beta\hat{e}_2} \bigg)
+ H.c. \bigg\} \bigg],
\end{eqnarray}
where $\hat{e}_j=a \hat{j}$ with $\hat{j}$ as the unit vector in along $j=x,y$, $\Psi^\top_{\vec{r}}=(c_{\vec{r},+},c_{\vec{r},-})$, and $c_{\vec{r},\tau}$ is the fermionic annihilation operator at $\vec{r}$ with parity $\tau=\pm$.

\emph{Chern number}.~Topological properties of this model can be tracked by computing the first Chern number of the filled valence band, for example, given by~\cite{TKNN}
\allowdisplaybreaks[4] 
\begin{equation}~\label{eq:chernnumber}
C=-\int_{\rm BZ} \dfrac{d^{2}{\vec k}}{4\pi} \:\: \big[ \partial_{k_x} \hat{\vec{d}}(\vec{k}) \times \partial_{k_y} \hat{\vec{d}}(\vec{k}) \big] \cdot \hat{\vec{d}}(\vec{k}),
\end{equation}
where $\hat{\vec{d}}(\vec{k})=\vec{d}(\vec{k})/|\vec{d}(\vec{k})|$. The integral is restricted within the first Brillouin zone (BZ). The resulting phase diagram is shown in Fig.~\ref{fig:Fig1}(c). It accommodates Chern insulators with the band inversion at the $\Gamma$ and ${\rm M}$ points with $C=-1$ and $+1$, respectively, named $\Gamma$ and ${\rm M}$ phases. They represent strong TIs with the band inversion at an odd number of points in the BZ. The latter is translationally active, as the ${\rm M}$ point results from the translational symmetries of the SL. Due to longer range hopping ($t_2$ and $\tilde{B}$), the above model also supports a Chern insulator with the band inversion simultaneously around the ${\rm X}$ and ${\rm Y}$ points of the BZ [Fig.~\ref{fig:Fig1}(d)], connected by four-fold rotations, with $C=-2$, representing a crystalline TI. The normal insulator (NI) has $C=0$.

\begin{table*}[t!]
\centering
\begin{tabular}{|ccccc|ccccc|}
\hline
\multicolumn{5}{|c|}{Symmetry-adapted terms in the Brillouin zone (BZ)}  & \multicolumn{5}{c|}{Symmetry-adapted terms in the real space (RS)}        
      \\ \hline \hline
\multicolumn{1}{|c|}{Function} & \multicolumn{1}{c|}{$R^{\rm BZ}_{\pi/2}$} & \multicolumn{1}{c|}{$R^{\rm BZ}_x$} & \multicolumn{1}{c|}{$R^{\rm BZ}_y$} & \multicolumn{1}{c|}{$\mathcal K$} & \multicolumn{1}{c|}{Function} & \multicolumn{1}{c|}{$R^{\rm RS}_{\pi/2}$} & \multicolumn{1}{c|}{$R^{\rm RS}_x$} & \multicolumn{1}{c|}{$R^{\rm RS}_y$} & \multicolumn{1}{c|}{$\mathcal K$} \\ \hline
\multicolumn{1}{|c|}{$S_x $, $S_x C_y$} & \multicolumn{1}{c|}{$-S_y $,$-S_y C_x$} & \multicolumn{1}{c|}{$S_x$,$S_x C_y$} & \multicolumn{1}{c|}{$-S_x$,$-S_x C_y$} & -,- & \multicolumn{1}{c|}{$i \cos \phi_{jk}$} & \multicolumn{1}{c|}{$-i \sin \phi_{jk}$} & \multicolumn{1}{c|}{$i \cos \phi_{jk}$} & \multicolumn{1}{c|}{$-i\cos \phi_{jk}$} & - \\ \hline
\multicolumn{1}{|c|}{$S_y$, $S_y C_x$} & \multicolumn{1}{c|}{$S_x$, $S_x C_y$} & \multicolumn{1}{c|}{$-S_y$, $-S_y C_x$} & \multicolumn{1}{c|}{$S_y$, $S_y C_x$} & -,- & \multicolumn{1}{c|}{$i \sin \phi_{jk}$} & \multicolumn{1}{c|}{$i \cos \phi_{jk}$} & \multicolumn{1}{c|}{$-i \sin \phi_{jk}$} & \multicolumn{1}{c|}{$i \sin \phi_{jk}$} & - \\ \hline
\multicolumn{1}{|c|}{$C_x + C_y$, $C_x C_y$} & \multicolumn{1}{c|}{$C_x + C_y$, $C_x C_y$} & \multicolumn{1}{c|}{$C_x + C_y$, $C_x C_y$} & \multicolumn{1}{c|}{$C_x + C_y$, $C_x C_y$} & +,+ & \multicolumn{1}{c|}{$C$} & \multicolumn{1}{c|}{$C$} & \multicolumn{1}{c|}{$C$} & \multicolumn{1}{c|}{$C$} & + \\ \hline
\multicolumn{1}{|c|}{$\bar{M}$} & \multicolumn{1}{c|}{$\bar{M}$} & \multicolumn{1}{c|}{$\bar{M}$} & \multicolumn{1}{c|}{$\bar{M}$} & + & \multicolumn{1}{c|}{$\bar{M}$} & \multicolumn{1}{c|}{$\bar{M}$} & \multicolumn{1}{c|}{$\bar{M}$} & \multicolumn{1}{c|}{$\bar{M}$} & + \\ \hline
\end{tabular}
\caption{Symmetry analyses of various terms appearing in the momentum space [Eqs.~\eqref{eq:QWZ} and~\eqref{eq:dvec}] and real space [Eq.~\eqref{eq:Method1}] Hamiltonian is shown in the first and last five columns, respectively. Here, $S_j \equiv \sin(k_j a)$ and $C_j \equiv \cos(k_j a)$ for $j=x,y$, $\bar{M}=M-4B-4\tilde{B}$ and $C$ are real constants, ${\boldsymbol k}=(k_x,k_y)$ is the momentum, and $\phi_{jk}$ is the azimuthal angle between sites $j$ and $k$, measured about the horizontal direction. Terms transforming identically under \emph{all} symmetry transformations appear in the same row. Here, $R^{\rm BZ}_{\pi/2}$ represents a $\pi/2$ rotation about the $z$ direction under which ${\boldsymbol k} \to (-k_y,k_x)$, while $R^{\rm BZ}_x$ [$R^{\rm BZ}_y$] corresponds to reflection about the $x$ [$y$] axis under which ${\boldsymbol k} \to (k_x,-k_y)$ [${\boldsymbol k} \to (-k_x,k_y)$]. On the other hand, $R^{\rm RS}_{\pi/2}$ represents a $\pi/2$ rotation about the $z$ direction in the real space under which $\phi_{jk} \to \phi_{jk}+\pi/2$, and $R^{\rm RS}_x$ [$R^{\rm RS}_y$] corresponds to reflection about the $x$ [$y$] axis in the real space under which $\phi_{jk} \to 2 \pi-\phi_{jk}$ [$\phi_{jk} \to \pi-\phi_{jk}$]. We summarize the transformation of each term under the complex conjugation (${\mathcal K}$), with ${\mathcal K}{\boldsymbol k} \to -{\boldsymbol k}$. In the fifth and tenth columns $+$ $(-)$ corresponds to even (odd). 
}~\label{tab:symmetryanalysis}
\end{table*}

\emph{Bott index}.~We aim to harness these phases on the SCFLs, where the notion of a BZ becomes moot due to the absence of the translational symmetry. Thus, we bring a related topological invariant onto the stage, the Bott index (BI), computed from the Hilbert space of the associated real space Hamiltonian $H^{\rm gen, SL}_{\rm QWZ}$ on a SL, obtained via a Fourier transformation of $H^{\rm gen}_{\rm QWZ}$, satisfying $H^{\rm gen, SL}_{\rm QWZ} \ket{E} = E \ket{E}$. First, we define two diagonal matrices, ${\boldsymbol X}$ and ${\boldsymbol Y}$, with their respective matrix elements given by $X_{i,j}=x_i \delta_{i,j}$ and $Y_{i,j}=y_i \delta_{i,j}$, encoding the position $(x_i,y_i)$ of the $i$th site, from which we define two unitary matrices ${\boldsymbol U}_X=\exp(2 \pi i {\boldsymbol X})$ and ${\boldsymbol U}_Y=\exp(2 \pi i {\boldsymbol Y})$. Next, in terms of the projector onto the filled eigenstates of $H^{\rm gen, SL}_{\rm QWZ}$, up to the Fermi energy $E_F=0$, defined as ${\mathcal P}=\sum_{E<E_F} \ket{E}\bra{E}$, we compute~\cite{Bottindex}
\begin{equation}~\label{eq:bottindex}
{\rm BI}= \frac{1}{2 \pi} {\Im} \left( {\rm Tr} \left[ \ln \left( {\bf V}_X {\bf V}_Y {\bf V}^\dagger_X {\bf V}^\dagger_Y \right) \right]\right),
\end{equation}   
in systems with periodic boundary conditions (PBCs), where ${\bf V}_j= {\rm I}-{\mathcal P} + {\mathcal P} {\boldsymbol U}_j {\mathcal P}$ for $i=X,Y$, showing ${\rm BI} \equiv C$. Thus, BI yields an identical phase diagram as in Fig.~\ref{fig:Fig1}(c).

\emph{Method 1}.~In this method, also named ``method of symmetry", we replace each term appearing in $\vec{d}(\vec{k})$ [Eq.~\eqref{eq:dvec}], constituting the Bloch Hamiltonian [Eq.~\eqref{eq:QWZ}], by its symmetry analogous term in the real space, such that both transform identically under all the discrete symmetry operations, the four-fold rotation about the $z$ direction, and the reflections about $x$ and $y$ axes. See Table~\ref{tab:symmetryanalysis}. The resulting real space Hamiltonian on SCFLs then reads
\allowdisplaybreaks[4]
\begin{widetext}
\begin{eqnarray}~\label{eq:Method1}
H^{\rm gen, 1}_{\rm QWZ} &=& \sum_{j \neq k} \frac{\Theta(R_1 - r^{\rm PA}_{jk})}{2} \exp\left[ 1-\frac{r^{\rm PA}_{jk}}{r^{\rm PA}_0}\right]  c^\dagger_j \bigg[ -i t_1 \left( \tau_1 \cos \phi^{\rm PA}_{jk}  + \tau_2 \sin \phi^{\rm PA}_{jk} \right) + 2B \tau_3 \bigg] c_k 
+ \sum_{j \neq k} \frac{\Theta(R_2 - r^{\rm BD}_{jk})}{2} \nonumber \\
& \times & \exp\left[ 1-\frac{r^{\rm BD}_{jk}}{r^{\rm BD}_0}\right] c^\dagger_j \bigg[ -i \frac{t_2}{\sqrt{2}} \big( \tau_1 \sin \phi^{\rm BD}_{jk} + \tau_2 \cos \phi^{\rm BD}_{jk} \big) + 2 \tilde{B} \tau_3 \bigg] c_k
+ \sum_{j} c^\dagger_j \left[ M-4 B -4 \tilde{B} \right] \tau_3 c_j, 
\end{eqnarray}
\end{widetext}
where $r^{\alpha}_{jk}=|\vec{r}_j-\vec{r}_k|$ ($\phi^{\alpha}_{jk}$) is the distance (azimuthal angle) between the $j$th and $k$th sites, located at $\vec{r}_j$ and $\vec{r}_k$, respectively, placed along the principal axes ($\alpha={\rm PA}$) and body diagonals ($\alpha={\rm BD}$), $c^\top_j=\left( c_{+,j}, c_{-,j} \right)$ is a two-component spinor, and $c_{\tau,j}$ is the fermion annihilation operator with parity $\tau=\pm$ on the $j$th site. In this construction, $R_1$ ($R_2$) controls the range of exponentially decaying hopping along PA (BD) through the Heaviside step function $\Theta$. Throughout, we set $r^{\rm PA}_0=r^{\rm BD}_0=a$.

\begin{figure}[t!]
\includegraphics[width=0.95\linewidth]{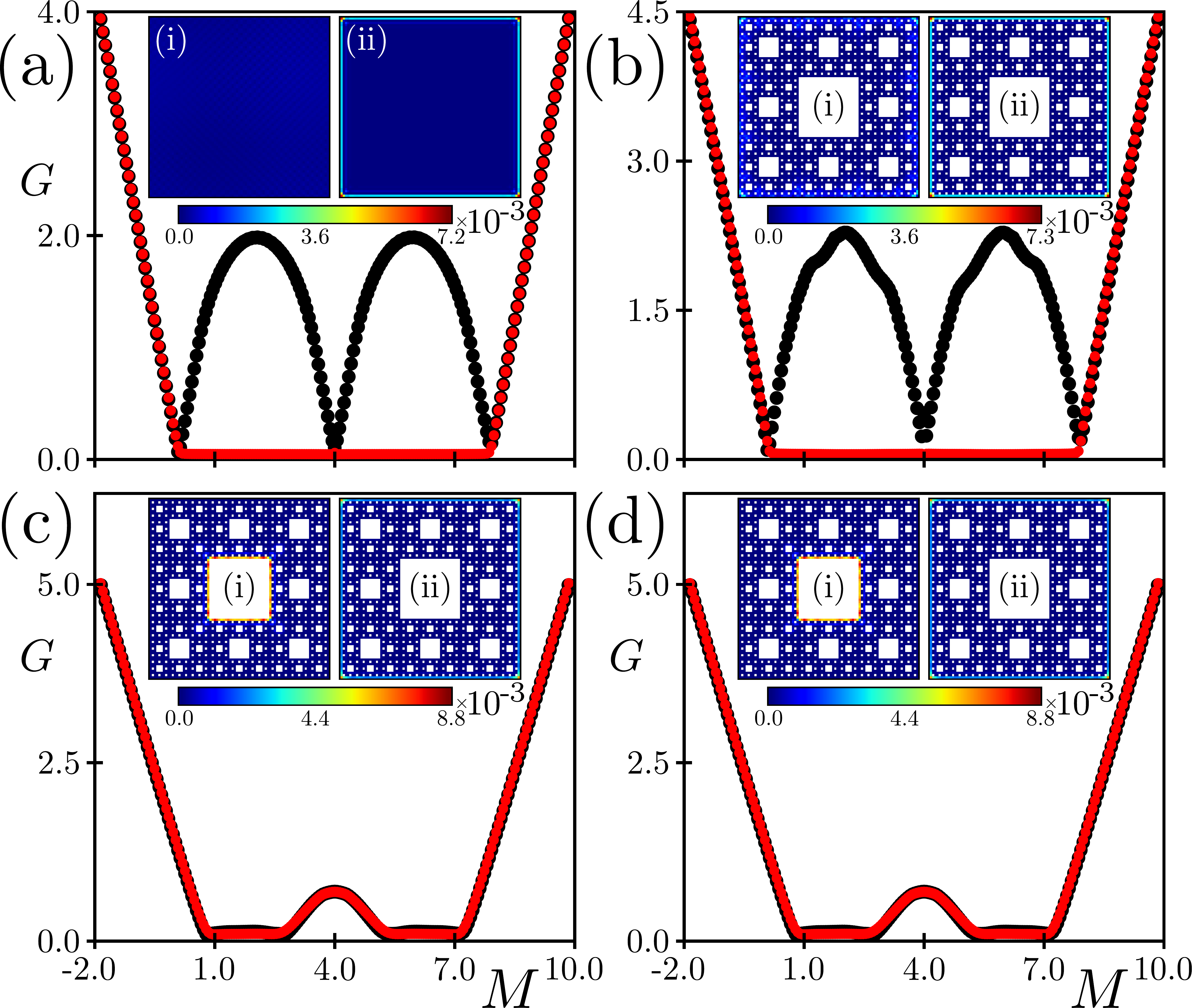}
\caption{Spectral gap between two closest-to-zero-energy modes, living on either side of it, for $t_1=B=1$, and $t_2=\tilde{B}=0$ on (a) an $L=81$ parent square lattice, and fourth generation Sierpi\'nski carpet fractal lattice, obtained by (b) method 3, (c) method 1, with $R_1=(1+\delta)a$, where $\delta \ll 1$ is a small positive number, and (d) method 2 with periodic (black dots) and open (red dots) boundary conditions. Insets show their local density of states for $M=6$, yielding ${\rm BI}=+1$ with (i) periodic and (ii) open boundary conditions. The spectral weight of inner edge modes, as shown in (ci) and (di), is dominantly localized around the central square hole, while that around the smaller square holes is negligibly small.  
}~\label{fig:Fig3}
\end{figure}

\emph{Method 2 and Method 3}.~Any tight-binding Hamiltonian on a SL ($H_{\rm SL}$) can be cast as a block matrix 
\allowdisplaybreaks[4]
\begin{eqnarray}~\label{eq:SLblockmatrix}
H_{\rm SL} = \left( \begin{array}{cc}
H_{\bullet \bullet} & H_{\bullet {\color{red} \bullet}} \\
H_{{\color{red} \bullet} \bullet} & H_{{\color{red}\bullet \bullet}}
\end{array}
\right),
\end{eqnarray}
where $H_{\bullet \bullet}$ ($H_{{\color{red}\bullet \bullet}}$) is the part of $H_{\rm SL}$ operating only on the black (red) colored sites [see Fig.~\ref{fig:Fig1}(b)], and $H_{\bullet {\color{red} \bullet}}$ and $H_{{\color{red} \bullet} \bullet}=H^\dagger_{\bullet {\color{red} \bullet}}$ capture the coupling between them. In method 2, also named ``Method of site elimination", the effective Hamiltonian for SCFLs is given by 
\allowdisplaybreaks[4]
\begin{equation}~\label{eq:Method2}
H^{\rm gen, 2}_{\rm QWZ}= H_{\bullet \bullet}\;.
\end{equation}
Here contributions from the red sites is completely ignored. In method 3, also named ``method of renormalization", the effective or renormalized Hamiltonian for SCFLs is constructed by integrating out the red colored sites of the parent SL, yielding
\allowdisplaybreaks[4]
\begin{equation}~\label{eq:Method3}
H^{\rm gen, 3}_{\rm QWZ}=
H_{\bullet \bullet}- 
H_{\bullet {\color{red}\bullet}} \: 
 H^{-1}_{{\color{red}\bullet \bullet}} \: 
H_{{\color{red}\bullet} \bullet},
\end{equation}  
assuming that $H^{-1}_{{\color{red}\bullet \bullet}}$ exists. This condition is satisfied as possible singularities (zero-energy modes
of $H_{{\color{red}\bullet \bullet}}$) are always isolated, and therefore can be regularized by taking a proper limiting procedure~\cite{blockmatrix}. This is so because the gap at the bulk or boundary nodal point scales as $\sim 1/L_R$, where $L_R$ is the linear system size, constituted by the red sites, with the nodal point pinned at zero energy only in the thermodynamic limit ($L_R \to \infty$). With all three general methodologies of constructing the Hamiltonian for SCFLs being staged, we now proceed to showcase the incarnation of all the TIs, accommodated by the SL Qi-Wu-Zhang model, on such systems. The results are summarized in Fig.~\ref{fig:Fig2}, which we discuss next. All the numerical codes are available on Zenodo~\cite{zenodo}.

\emph{Results}.~A SCFL is constructed from a parent SL in the following way. We divide a SL into $3 \times 3$ squares. Then we remove the central square. We repeat this procedure recursively for each of the eight remaining squares to obtain different generations ($g$). In the $g$th generation the total number of squares is $9^g$ and the total number of unremoved squares is $8^g$. Hence the SCFL has a fractal dimension $d_{\rm frac}=\ln(8^g)/\ln(\sqrt{9^g}) \approx 1.89$. The topological phases in the SCFLs of any $g$ can be identified from the BI by diagonalizing the corresponding real space Hamiltonian, shown in Eqs.~\eqref{eq:Method1},~\eqref{eq:Method2}, and~\eqref{eq:Method3}.

Irrespective of $g$, method 1 and method 2, resulting in the Hamiltonian in Eq.~\eqref{eq:Method1} and Eq.~\eqref{eq:Method2}, respectively, yield identical phase diagrams for SCFLs for various parameter values therein, when in the former setup we set $R_1=(1+\delta)a$ and $R_2=(\sqrt{2}+\delta)a$, where $\delta \ll 1$ is a small positive number. This observation assures the existence of various phases, appearing in the phase diagrams, in the thermodynamic limit. Specifically, Fig.~\ref{fig:Fig2}(a) displays Chern insulators with ${\rm BI}=-1$ and $+1$, identical to those for the $\Gamma$ and ${\rm M}$ phases, respectively. Figure~\ref{fig:Fig2}(b) shows the appearance of a Chern insulator with ${\rm BI}=-2$, as found in the `valley XY phase'. These phase diagrams are qualitatively similar to the one shown in Fig.~\ref{fig:Fig1}(c) for a SL, obtained in terms of the first Chern number and BI. Phase diagrams of a $g=3$ SCFL, obtained via method 3, are identical to those found in the parent SL of linear dimension $L=27$. For example, Fig.~\ref{fig:Fig2}(c) accommodates Chern insulators with ${\rm BI}=+1$ and $-1$, whereas Fig.~\ref{fig:Fig2}(d) shows a Chern insulator with ${\rm BI}=-2$. The phase diagrams in the $(M, \tilde{B})$ plane, obtained from three methods, are shown in the Supplemental Material~\cite{SM}.

Note that while the range of hopping in $H^{\rm gen, 1}_{\rm QWZ}$ [Eq.~\eqref{eq:Method1}] (obtained from method 1) can be tuned by $R_1$ and $R_2$, we chose their values such that only the nearest-neighbor and next-nearest-neighbor ones are operative. By contrast, $H^{\rm gen, 2}_{\rm QWZ}$ [Eq.~\eqref{eq:Method2}], directly obtained from $H^{\rm SL}_{\rm TB}$ [Eq.~\eqref{eq:realspaceTB}] after eliminating the sites falling outside the fractal lattice (method 2), contains only these two hopping processes. Consequently, these two methods yield identical phase diagrams, fostering the same TIs as the parent SL, with the ${\rm BI}=\pm 1, -2$ and a NI with ${\rm BI}=0$. However, the parameter regimes over which these phases are realized for $H^{\rm gen, 1/2}_{\rm QWZ}$ are different from the ones for the parent SL [Fig.~\ref{fig:Fig1}(c)], as the exact form of the corresponding real-space Hamiltonian are different. Finally, we note that, although $H^{\rm gen, 3}_{\rm QWZ}$ and $H_{\rm SL}$ are of different dimensions [compare Eqs.~\eqref{eq:Method3} and \eqref{eq:SLblockmatrix}], the former Hamiltonian is defined in terms of longer-ranged hopping with renormalized amplitudes, as it is obtained by integrating out red colored sites of the SL [Fig.~\ref{fig:Fig1}(b)], captured by the second term of Eq.~\eqref{eq:Method3}, falling outside SCFL. Therefore, even though $H_{\rm SL}$ is defined in terms of only nearest-neighbor and next-nearest-neighbor hopping amplitudes, the renormalized hopping range and amplitude of $H^{\rm gen, 3}_{\rm QWZ}$ ensure that it possesses identical topological properties of $H_{\rm SL}$, directly following the definition of the renormalization procedure. Consequently, the phase diagram obtained from method 3 is identical to the one for the parent SL in the entire parameter regime~\cite{SM}.

Topological phases with nontrivial and quantized BI support topological edge modes, manifesting the hallmark bulk-boundary correspondence. On a SL, the near-zero-energy topological edge modes are found only in systems with open boundary conditions (OBCs) [Fig.~\ref{fig:Fig3}(a)]. Due to the self-similarity symmetry, 2D fractal lattices harbor outer and inner edges. The topological Hamiltonian constructed in method 3 ($H^{\rm gen, 3}_{\rm QWZ}$), however, accommodates such near-zero-energy modes on SCFLs only with OBCs that are localized only near the outer edges [Fig.~\ref{fig:Fig3}(b)]. This Hamiltonian does not support any near-zero-energy modes close to the inner edges of SCFLs with PBCs. This is so because $H^{\rm gen, 3}_{\rm QWZ}$ [Eq.~\eqref{eq:Method3}] is constructed by systematically integrating out the red sites of the parent SL, thereby inheriting all the spectral properties of the parent crystal. By contrast, the Hamiltonian for the SCFLs, constructed from method 1 [Eq.~\eqref{eq:Method1}] and method 2 [Eq.~\eqref{eq:Method2}] support near-zero-energy modes close to their outer and inner edges in systems with OBCs and PBCs [Figs.~\ref{fig:Fig3}(c) and~\ref{fig:Fig3}(d)], respectively, as they are constructed by ignoring any influence of the red sites of the parent SL. Thus these two methods expose the inner boundaries of the self-similar fractal lattices. Although in Fig.~\ref{fig:Fig3}, we display these results for a Chern insulator with ${\rm BI}=+1$, we arrive at qualitatively similar results for those with ${\rm BI}=-1$ and $-2$ (not show explicitly). Finally, we note that SCFLs also support NIs with ${\rm BI}=0$, devoid of any near-zero-energy outer or inner edge modes.

\emph{Summary and discussions}.~Here we formulate three independent approaches to construct the effective real space Hamiltonian on fractal lattices starting from the Bloch Hamiltonian in their parent crystals to harness different classes of TIs therein, namely the strong and crystalline ones. We believe that none of these methods can describe the effective Hamiltonian on fractal lattices in real materials in full accuracy. Nonetheless, given that all three methods permit strong and crystalline TIs, with method 2 and method 3 corresponding to two extreme limits, and feature the signature bulk-boundary correspondence, it is highly conceivable that all these phases can also be found in real fractal lattices, nowadays realizable in designer electronic~\cite{frac:exp1, frac:exp2} and molecular~\cite{frac:exp3} systems as well as in classical metamaterials~\cite{frac:exp4, frac:exp5, frac:exp6}.

Our proposed methodologies can be employed to capture topological phases on any fractal lattice belonging to any Altland-Zirnbauer symmetry class and any crystalline group in any dimension (such as the 2D hexaflake and three-dimensional Menger sponge), as long as there exists a parent topological crystal (honeycomb and cubic, respectively). Our theoretical framework, therefore, opens promising possibilities to realize (both theoretically and experimentally) a vast variety of topological phases of matter on the rich landscape of fractal lattices, going beyond the existing studies of specific topological models on specific fractal lattices~\cite{frac:th1, frac:th2, frac:th3, frac:th4, frac:th5, frac:th6, frac:th7, frac:th8, frac:th9, frac:th10, frac:th11, frac:th12, frac:th13, frac:th14, frac:th15, frac:th16, frac:th17}. We also note that there is no sharp topological bulk gap when effective Hamiltonian are constructed from methods 1 and 2 [Fig.~\ref{fig:Fig3}] due to in-gap modes localized near the inner boundaries of SCFLs, raising a question of fundamental and practical importance regarding the stability of TIs on fractal lattices in the inevitable presence of disorder. These fascinating questions are reserved for systematic future investigations. For the model considered here, we show that all the TIs are stable against sufficiently weak disorder, while a normal or trivial insulator with ${\rm BI}=0$ appears in strong disorder regime~\cite{SM}. 

\emph{Acknowledgments}.~This work was supported by the NSF CAREER Grant No.\ DMR-2238679 of B.R. We are thankful to Vladimir Juri\v{c}i\'c and Sanjib Kumar Das for a critical reading of the manuscript.


\end{document}